\documentclass[12pt]{article}

\advance\voffset by -2.0cm \advance\hoffset by -1.25cm
\usepackage{amsmath}
\usepackage{amssymb}
\usepackage{amsthm}
\usepackage{amsfonts}

\theoremstyle{definition}

\newcommand{\CC}{\mathbb{C}} 
\newcommand{\RR}{\mathbb{R}} 

\newcommand{\be}{\begin{equation}}
\newcommand{\ee}{\end{equation}}

\newlength{\oldcolsep}

\numberwithin{equation}{section}

\begin{document}

\title{Closed Form of the Baker-Campbell-Hausdorff\\ Formula for the Generators of Semisimple\\ Complex Lie Algebras}

\author{Marco Matone}\date{}

\maketitle

\begin{center} Dipartimento di Fisica e Astronomia ``G. Galilei'' \\
 Istituto
Nazionale di Fisica Nucleare \\
Universit\`a di Padova, Via Marzolo, 8-35131 Padova,
Italy\end{center}

\begin{abstract}
\noindent
Recently it has been introduced an algorithm Baker-Campbell-Hausdorff (BCH) formula, which extends the Van-Brunt and Visser recent results, leading to new closed forms of BCH formula.
More recently, it has been shown that there are {\it 13 types} of such commutator algebras. We show, by providing the explicit solutions, that these include
the generators of the semisimple complex Lie algebras.
More precisely, for any pair, $X$, $Y$ of the Cartan-Weyl basis, we find $W$, linear combination of $X$, $Y$, such that
$$
\exp(X) \exp(Y)=\exp(W)
$$
The derivation of such closed forms follows, in part, by using the above mentioned recent results. The complete derivation is provided by considering the structure of of the root system.
Furthermore, if $X$, $Y$ and $Z$ are three generators of the Cartan-Weyl basis, we find, for a wide class of cases, $W$, linear combination of $X$, $Y$ and $Z$, such that
$$
\exp(X) \exp(Y) \exp(Z)=\exp(W)
$$
It turns out that the relevant commutator algebras are {\it type 1c-i}, {\it type 4} and {\it type 5}. A key result concerns an iterative application of the algorithm leading
to relevant extensions of the cases admitting closed forms of the BCH formula. Here we provide the main steps of such an iteration that will be developed in a forthcoming paper.

\end{abstract}

\newpage

\section{Introduction}

The Baker-Campbell-Hausdorff (BCH) formula
is of considerable interest in Quantum Mechanics, Quantum Field Theory, Conformal Field Theory, Statistical Mechanics, Quantum Computing, Optics, etc.  (see, for example \cite{DiFrancesco:1997nk}-\cite{Sornborger:1999rt}). It is then clear that it is of interest to find cases when the BCH formula admits a closed form. Such a formula expresses
\be
Z = \ln(e^X e^Y) \ ,
\ee
in terms of an infinite series of commutators of $X$ and $Y$.

\noindent
In \cite{Matone:2015wxa}
it has been introduced a simple algorithm leading, for a wide class of cases, including some Virasoro subalgebras, to closed forms of the BCH formula.
Such an algorithm  exploits the associativity of the BCH formula and implement the Jacobi identity.
 In
 \cite{Matone:2015xaa}
 it has been shown that there are
{\it 13
types} of commutator algebras admitting such simplified versions of the BCH formula. In \cite{Matone:2015hja} the closed form of the BCH formula has been used to covariantize
the conformal group.
The main points of the algorithm are the following. First, following the Van-Brunt and Visser remarkable result, consider
two elements, $X$ and $Y$, of an associative
algebra, with commutator
\begin{equation}
[X,Y]=uX+vY+cI \ ,
\label{AAAbeoftheform}\end{equation}
with $u$, $v$ and $c$, being $c$-numbers and $I$ a central element. Van-Brunt and Visser proved that \cite{Van-Brunt:2015ala}
(see also \cite{Van-Brunt:2015bza,Van-Brunt:2015hsa})
\begin{equation}
\exp(X) \exp(Y)= \exp({X+Y+f(u,v)[X,Y]}) \ ,
\label{bbbvvv}\end{equation}
where $f(u,v)$ is the symmetric function
\be
f(u,v)={(u-v)e^{u+v}-(ue^u-ve^v)\over uv(e^u-e^v)} \ .
\ee
In \cite{Matone:2015wxa} it has been considered the BCH problem of finding the closed form of $W$ in
\be
\exp(X)\exp(Y)\exp(Z)=\exp(W) \ .
\label{BCHproblem}\ee
The first step of the algorithm is to consider the decomposition
\begin{equation}
\exp(X) \exp(Y) \exp(Z)= \exp(X)\exp({\alpha Y}) \exp({\beta Y}) \exp(Z) \ ,
\label{decomposizione}\end{equation}
$\alpha+\beta=1$.
If
\begin{equation}
[X,Y]=uX+vY+cI \ , \qquad [Y,Z]=wY+zZ+dI \ ,
\label{riecco}\end{equation}
then, by (\ref{bbbvvv}) and (\ref{decomposizione}),
\be
\exp(X) \exp(Y) \exp(Z)= \exp({\tilde X})\exp({\tilde Y}) \ ,
\label{llss}\ee
where $\tilde X:=\ln(e^Xe^{\alpha Y})$ and $\tilde Y :=\ln(e^{\beta Y}e^Z)$. By (\ref{bbbvvv}) and noticing that $[X,\alpha Y]=(\alpha u)X+v(\alpha Y)+(\alpha c)I$ and $[\beta Y,Z]=w(\beta Y)+(\beta z)Z+(\beta d)I$
\begin{align}
\tilde X &=X+\alpha Y+ f(\alpha u,v)[X,\alpha Y]  \ , \cr
\tilde Y &=\beta Y + Z + f(z,\beta w)[\beta Y,Z] \ ,
\label{colon0prima}\end{align}
that we rewrite in the form
\begin{align}
\tilde X &:=g_{\alpha}(u,v)X+h_{\alpha}(u,v)Y+l_{\alpha}(u,v)cI \ , \cr
\tilde Y &:=h_{\beta}(z,w)Y+g_{\beta}(z,w)Z+l_{\beta}(z,w)dI \ ,
\label{colon0}\end{align}
with
$$
g_\alpha(u,v):=1+\alpha uf(\alpha u, v) \ , \quad h_{\alpha}(u,v):=\alpha(1+v f(\alpha u, v))\ , \quad
l_{\alpha}(u,v):=\alpha f(\alpha u, v) \ .
$$
Next, imposing that $\exp (\tilde X)\exp (\tilde Y)$ satisfies the condition to apply (\ref{bbbvvv}),
that is
\begin{equation}
[\tilde X,\tilde Y]=\tilde u \tilde X+\tilde v \tilde Y+\tilde c I \ ,
\label{labella}\end{equation}
we get the solution of (\ref{BCHproblem}) \cite{Matone:2015wxa}
\begin{equation}
\exp(X) \exp(Y) \exp(Z)= \exp({\tilde X+\tilde Y+f(\tilde u,\tilde v)[\tilde X,\tilde Y]}) \ .
\label{llasol}\end{equation}

\noindent Note that, by (\ref{riecco}), (\ref{colon0}) and (\ref{labella}), it follows that $[X,Z]$ cannot be a linear combination of other terms in addition to
$X$, $Y$, $Z$ and $I$, that is
\begin{equation}
[X,Z]=m X+nY+pZ+e I \ ,
\label{richiamare}\end{equation}
which is consistent with the Jacobi identity
\be
[X,[Y,Z]]+[Y,[Z,X]]+[Z,[X,Y]]=0 \ .
\label{JIGOOD}\ee
The condition (\ref{labella})
fixes $\alpha$, $\tilde c$, $\tilde u$ and $\tilde v$, namely
$$
\tilde c=(h_\beta (z,w)-g_\beta(z,w)l_\alpha(u,v) m) c +(h_\alpha(u,v)-g_\alpha(u,v) l_\beta(z,w) p)d+g_\alpha(u,v)g_\beta(z,w) e \ ,
$$
\begin{align}
\tilde u&=h_{\beta}(z,w)u+g_{\beta}(z,w)m \ ,\cr
\tilde v&=g_{\alpha}(u,v)p+h_{\alpha}(u,v)z \ ,
\label{utildevtilde}\end{align}
with $\alpha$ solution of the equation
\begin{equation}
h_{\alpha}(u,v) [h_{\beta}(z,w)(u+z)+g_{\beta}(z,w)(m-w)]+g_{\alpha}(u,v)
[h_{\beta}(z,w)(p-v)-g_{\beta}(z,w)n]=0 \ .
\label{fonda}\end{equation}
Inserting (\ref{riecco}) and (\ref{richiamare}) in the Jacobi identity leads to the following linear system for $e$, $m$, $n$ and $p$
\begin{align}
uw+mz & =0 \ ,  \cr
vm-wp + n(z-u) & = 0 \ ,  \cr
pu+zv & = 0 \ ,  \cr
c(w+m)+e(z-u)-d(p+v)&=0 \ .
 \label{AsystemINTRO}\end{align}
Such a system has thirteen different solutions, reported in the table, depending on the values and algebraic relations of $c$, $d$, $u$, $v$, $w$ and $z$.
Note that the case $cw=dv=0$ corresponds to five different conditions (see \cite{Matone:2015xaa}).
\\

\begin{center}
 \centerline{\it The thirteen cases of the Jacobi identity}
 \vspace{.3cm}
\begin{tabular}{|c|c|c|c|c|}
 \hline {\it 1} & {\it 2} & {\it 3} & {\it 4} &  {\it 5} \\
 \hline $u=z=0$ & $u=0, \; z\neq0$ & $u\neq0,\; z=0$ & $u=z\neq0$ &  $u\neq z,\; uz\neq0$ \\ \hline   $cw\neq dv$ & $w=0$ & $v=0$ & {} & \\
$cw=dv\neq0$ &  $w\neq0$ &  $v\neq 0$ & {} & \\
$cw=dv=0$ & {} & {} & {} & {}  \\ \hline
\end{tabular}
\end{center}

\vspace{.6cm}

\noindent Consider the symbol \cite{Matone:2015xaa}
\\
$$\Big[case\Big| Jacobi \; constraints\Big|parameters \; of\;  [X,Z]\; unfixed\Big]_{D}  $$\\
The first slot specifies under which conditions the linear system (\ref{AsystemINTRO}) is solved. This classifies the
{\it types} of commutator algebras.
The constraints are the ones on the commutator parameters that follow by the Jacobi identity. The third slot
reports which ones, between the parameters $m$, $n$, $p$ and $e$ in  $[X,Z]$, remains unfixed. $D$ is the number
of the commutator parameters unfixed by the Jacobi identity.

\noindent
In this paper we explicitly show that the above algorithm leads to closed forms for the BCH formula in the case of semisimple complex Lie algebras.
In particular, according to the above classification the commutator algebras, we will see that, in several cases, the commutator algebras associated to the BCH problem for semisimple complex Lie algebras
corresponds to the {\it type 1c-i}, {\it type 4} and {\it type 5}.
This implies that if $X$, $Y$ and $Z$ are three generators of the Cartan-Weyl basis, then, for a wide class of cases, $W$, defined by
(\ref{BCHproblem}) is explicitly expressed as a linear combination of $X$, $Y$ and $Z$.

\noindent In the last section we derive, by iteration, a basic generalization of the algorithm introduced in \cite{Matone:2015wxa}.
This provides important extensions of the cases for which the BCH formula admits a closed form.

\noindent Before discussing the case of arbitrary semisimple complex Lie algebras, we consider the BCH problem for ${\rm sl}_3(\CC)$.

\section{BCH formulas for the generators of ${\rm sl}_3(\CC)$}

 \noindent Consider the commutators for ${\rm sl}_3(\CC)$ (see, for example, \cite{Fuchs:1997jv})
$$
\begin{array}{lll} [H^1,H^2]=0 \ , &  [E_+^1,E_-^1]=H^1 \ , & [E_+^2,E_-^2]=H^2 \ , \cr\cr
[H^1,E_\pm^1]=\pm2 E_\pm^1 \ , & [H^1,E_\pm^2]=\mp E_\pm^2 \ , & [H^1,E_\pm^\theta]=\pm E_\pm^\theta \ , \cr\cr
[H^2,E_\pm^1]=\mp E_\pm^1 \ , &  [H^2,E^2_\pm]=\pm 2E_\pm^2 \ , & [H^2,E_\pm^\theta]=\pm E_\pm^\theta \ , \cr\cr
[E_\pm^1,E_\pm^2]=\pm E_\pm^\theta \ , & [E_\pm^1, E_\mp^\theta]=\mp E_\mp^2 \ , & [E_\pm^2,E_\mp^\theta]=\pm E_\mp^1 \ , \cr\cr
 [E_+^\theta,E_-^\theta]=H^1+H^2 \ , &  &
 \end{array}
 $$
with the remaining commutators vanishing
 $$
 [E_\pm^1,E_\mp^2]=[E_\pm^1,E_\pm^\theta]=[E_\pm^2,E_\pm^\theta]=0 \ .
$$
\noindent In the following we will consider the BCH problem of finding the closed form of $W$ in
\be
\exp(X)\exp(Z)=\exp(W) \ ,
\ee
with $X$ and $Z$ all the possible generators of ${\rm sl}_3(\CC)$.

\noindent Let us start with ${E_+^1}$, ${E_+^2}$, and ${E_-^1}$, ${E_-^2}$.
Since
\be
[E_+^1,[E_+^1,E_+^2]]=0=[E_+^2, [E_+^1,E_+^2]] \ ,
\ee
we have
\be
\exp({E_+^1})\exp({E_+^2})=\exp({E_+^1+E_+^2+E_+^\theta/2}) \ .
\ee
Similarly
\be
\exp({E_-^1})\exp({E_-^2})=\exp({E_-^1+E_-^2-E_-^\theta/2}) \ .
\ee
Next, consider $E_+^1$ and $E_-^\theta$. Also in this case, since
\be
[E_+^1,[E_+^1,E_-^\theta]]=0=[E_-^\theta,[E_+^1,E_-^\theta]] \ ,
\ee
we have
\be
\exp(E_+^1)\exp(E_-^\theta)=\exp(E_+^1+E_-^\theta-E^2_-/2) \ ,
\ee
and
\be
\exp(E_-^1)\exp(E_+^\theta)=\exp(E_-^1+E_+^\theta+E^2_+/2) \ .
\ee
Similar relations hold for $\exp(E_+^2)\exp(E_-^\theta)$ and $\exp(E_-^2)\exp(E_+^\theta)$.
Next, consider
\be
\exp(E_\pm^j)\exp(H^k) \ ,
\ee
$j,k=1,2$. All these are of the same kind, so that we consider only $j=k=1$. By (\ref{bbbvvv})
\be
\exp(E_+^1)\exp(H^1)=\exp\Big({2\over e^2-1}E_+^1+H^1\Big) \ .
\ee
Let us now consider the nontrivial cases
\be
\exp(E_+^k)\exp(E_-^k) \ ,
\label{kappauno}\ee
$k=1,2$, and
\be
\exp(E_-^\theta)\exp(E_+^\theta) \ .
\ee
We focus on (\ref{kappauno}) with $k=1$, the case
$k=2$ is equivalent.
This is a particular case of the {\it type 4} commutator algebras \cite{Matone:2015xaa},
and has been worked out, as an example, in \cite{Matone:2015wxa}
for ${\rm sl}_2(\RR)$
\begin{equation}
[L_m,L_n]=(n-m)L_{m+n} \ ,
\end{equation}
$m,n=-1,0,1$.
A straightforward application of the algorithm leads to  \cite{Matone:2015wxa}
\begin{align}
\exp({\lambda_{-1}L_{-1}})  &  \exp({\lambda_0L_0})
\exp({\lambda_1L_1})= \cr\cr
& \exp\Big\{{\lambda_+-\lambda_-\over e^{-\lambda_-}-e^{-\lambda_+}}\Big[\lambda_{-1} L_{-1}+\Big(2-e^{-\lambda_+}-e^{-\lambda_-}\Big)L_0+\lambda_1 L_1\Big]\Big\} \ ,
\label{lasoluzionee}\end{align}
where $\lambda_{-1},\lambda_0,\lambda_1\in\CC$, $\lambda_-+\lambda_+=\lambda_0$, and
\be
e^{-\lambda_{\pm}}={1+e^{-\lambda_0}-\lambda_{-1}\lambda_1\pm \sqrt{(1+e^{-\lambda_0}-\lambda_{-1}\lambda_1)^2-4e^{-\lambda_0}}\over2} \ .
\ee
Let us explicitly write down the case of $\exp({E_+^1})\exp({E_-^1})$. Setting $\lambda_0=0$, corresponding to $\lambda_-=-\lambda_+$, in Eq.(\ref{lasoluzionee})
and using the identification
\be
L_{-1}=-E_+^1 \ , \qquad L_0=-{H^1\over2} \ , \qquad L_1=E_-^1 \ ,
\ee
yields
\be
\exp({E_+^1})\exp({E_-^1})=\exp\Big[{2\over\sqrt 5}\ln {3+\sqrt 5\over 2}\Big(E_+^1+{H^1\over2}+E_-^1\Big)\Big] \ .
\label{originario}\ee
An identical expression holds in the case of ${E_-^2}$, ${E_+^2}$, one has just to replace $H^1$ by $H^2$. The same happens for $\exp(E_-^\theta)\exp(E_+^\theta)$. Actually, setting
$$
H:= H^1+H^2 \ ,
$$
we see that the relevant algebra to treat $\exp(E_-^\theta)\exp(E_+^\theta)$ is again ${\rm sl}(2,\RR)$
\be
[H,E_\pm^\theta]=\pm 2 E_\pm^\theta \ , \qquad [E_+^\theta,E_-^\theta]=H \ .
\ee
Therefore, $\exp(E_-^\theta)\exp(E_+^\theta)$ is again given by (\ref{originario}) by obvious substitutions.

\section{BCH formulas for the generators of semisimple complex Lie algebras}

\noindent Let $\Phi(\mathfrak g)$ be the root system of a semisimple complex Lie algebra $\mathfrak g$. Denote by $(\alpha,\beta)$ the standard non-degenerate inner product on the dual of the
Cartan subalgebra ${\mathfrak g}_0^*$, and
$$
\alpha^\vee:={2\alpha\over (\alpha,\alpha)} \ .
$$
For any $\mathfrak g$, consider its Cartan-Weyl basis
\begin{align}
[H^\alpha, H^\beta]&=0 \ , \cr\cr
[H^\alpha,E^\beta]&=(\alpha^\vee,\beta) E^\beta \ , \cr\cr
[E^\alpha,E^{-\alpha}]&=H^\alpha \ , \cr\cr
[E^\alpha, E^{\beta}]&=e_{\alpha\beta}E^{\alpha+\beta} \ , \qquad \alpha\neq-\beta \ ,
\label{lalgebra}\end{align}
where $e_{\alpha\beta}$, defined for any $\alpha\neq-\beta$, is non-zero if $\alpha+\beta\in \Phi(\mathfrak g)$, and zero otherwise.
In the following, for any $X$, $Y$ and $Z$, identified with any triple of the Cartan-Weyl generators $H^\alpha$ and $E^\beta$,
we consider the problem of finding the closed form of $W$ defined by (\ref{BCHproblem}). f
We will explicitly show that, for a wide class of cases, such closed forms are the ones classified in
\cite{Matone:2015xaa}.
In particular, the relevant commutator algebras
correspond to the {\it type 1c-i}, {\it type 4}  and {\it type 5} commutator algebras, leading to the closed forms of the BCH formula introduced in  \cite{Matone:2015wxa}.

\noindent
For any $a\in \CC$, set
\be
s(a):= {\sinh (a/2)\over a/2} \ .
\ee
The {\it type 1c-i} commutator algebras corresponds to the case
$$
c=d=u=z=0 \ ,
$$
denoted in \cite{Matone:2015xaa}
by $(v,w)$. It turns out that

\vspace{.8cm}

\noindent\underline{\it Type 1c-i.}

{$$\Big[(v,w) \Big| p={mv\over w}\Big|e,m,n\Big]_5  $$}
$$
[X,Y]=vY \qquad [Y,Z]=wY  \qquad [X,Z]=mX+nY+{mv\over w}Z+eI
$$

\begin{align}
\alpha & = {ns(v)ws(w)-e^{w\over2}vs(v)(m-w)\over (m-w)(w-v)s(w-v)} \nonumber \cr
\tilde u & = m \nonumber \cr
\tilde v & = {mv\over w} \nonumber \cr
\tilde c & =  e \nonumber
\end{align}

\noindent
The {\it type 4} corresponds to the case $u=z\neq0$, namely

\vspace{.8cm}

\noindent\underline{\it Type 4.}

{$$\Big[u=z\neq 0\Big|m=-w, \; p=-v\Big|e, n\Big]_{8}  $$}
$$
[X,Y]=uX+vY+cI  \qquad
[Y,Z]=wY+zZ+dI \qquad
[X,Z]=-wX+nY-vZ+eI
$$
$x^u$, where $x:= e^\alpha$, satisfies the equation
$$
x^{2u}+x^u\Big({nu\over2}s(v)s(w)e^{u+v-w\over2}-e^u-e^v+e^{u+v}-e^{u+v-w}\Big)+e^{u+v-w}=0
$$
Therefore
\begin{equation}
x_\pm^u = {-b\pm \sqrt{b^2-4e^{u+v-w}}\over2}
\label{laxuX}\end{equation}
where
$$
b:={nu\over2}s(v)s(w)e^{u+v-w\over2}-e^u-e^v+e^{u+v}-e^{u+v-w}
$$
\begin{align}
\tilde u & =\beta u-w  \cr
\tilde v & =\alpha u -v \cr
\tilde c & = \Big(e+{cw+dv\over u}\Big){ e^{u\over2}s(v)s(w)\over s(v-\alpha u)s(w-\beta u)}- {cw+dv\over u} +\beta c+\alpha d
\label{poffarre}\end{align}

\vspace{.8cm}

\noindent
Let us first consider the case $\exp(\lambda_\alpha H^\alpha) \exp(\mu_\beta E^\beta)$. This
corresponds to (\ref{bbbvvv}), so that
\begin{equation}
\exp(\lambda_\alpha H^\alpha) \exp(\mu_\beta E^\beta)= \exp\Big( \lambda_\alpha H^\alpha+{\lambda_\alpha\mu_\beta (\alpha^\vee,\beta)
\over 1-e^{-\lambda_\alpha(\alpha^\vee,\beta)}} E^\beta\Big) \ .
\label{lasemplice}\end{equation}
Let us consider
\be
\exp(\lambda_\alpha H^\alpha) \exp(\mu_\beta E^\beta)
\exp(\lambda_\gamma H^\gamma) \ .
\ee
This corresponds to the {\it type 1c-i} commutator algebras, where the only non-vanishing commutator parameters are
\be
v=\lambda_\alpha (\alpha^\vee,\beta) \ , \qquad w=-\lambda_\gamma(\gamma^\vee,\beta) \ .
\ee
Therefore,
\begin{align}
\alpha & = {1-e^{\lambda_\gamma(\gamma^\vee,\beta)}\over e^{-\lambda_\alpha(\alpha^\vee,\beta)}-e^{\lambda_\gamma(\gamma^\vee,\beta)}} \ , \cr
\tilde u & = 0 \ , \cr
\tilde v & = 0 \ ,  \cr
\tilde c & =  0 \ .
\end{align}
This leads to
\begin{align}
\exp(\lambda_\alpha H^\alpha) &\exp(\mu_\beta E^\beta)
\exp(\lambda_\gamma H^\gamma) = \cr\cr
&\exp\bigg[\lambda_\alpha H^\alpha+\lambda_\gamma H^\gamma+{\mu_\beta(\lambda_\alpha(\alpha^\vee,\beta)+\lambda_\gamma(\gamma^\vee,\beta))\over
e^{\lambda_\gamma(\gamma^\vee,\beta)}-e^{-\lambda_\alpha(\alpha^\vee,\beta)}}E^\beta\bigg]\ .
\end{align}
\\
Let us  now consider the BCH problem
$\exp(X)\exp(Y)=\exp(W)$,
with $X=\mu_\alpha E^\alpha$, $Y=\mu_\beta E^\beta$, $\alpha\neq-\beta$, $\alpha+\beta\in \Phi(\mathfrak g)$.
In this case it is convenient to use the expansion
\be
W=X+Y+{1\over2}[X,Y]+{1\over 12}([X,[X,Y]]+[Y,[Y,X]])-{1\over24}[Y,[X,[X,Y]]]+ \ldots \ .
\label{bchexpansion}\ee
Note that
\be
[E^\beta,[E^\alpha,[E^\alpha,E^\beta]] \ ,
\label{zerooo}\ee
would be proportional to $E^{2(\alpha+\beta)}$. On the other hand, since $\alpha+\beta$ is a root,
it follows that $2(\alpha+\beta)\notin \Phi(\mathfrak g)$. So that (\ref{zerooo}) vanishes. In order to consider the higher order terms in (\ref{bchexpansion}), we first recall
 that the length of root strings is at most five. Here we do not consider the special cases in which either $2\alpha+3\beta$ or $3\alpha+2\beta$ are roots, so that we specialize to the cases
 in which the length of root strings is at most four.
 Denote by $n_X$ ($n_Y$) the number of $X$ ($Y$) appearing in each term of the expansion (\ref{bchexpansion}).
 Then observe that all the terms in the dots of (\ref{bchexpansion}) correspond to either $n_X\geq4$ and $n_Y\geq1$, or  $n_Y\geq4$ and $n_X\geq1$, or  $n_X\geq2$ and $n_Y\geq2$.
Therefore the terms in (\ref{bchexpansion}) which are next to (\ref{zerooo}) are proportional to either
$E^{4\alpha+n\beta}$, $n\geq 1$, or $E^{n\alpha+4\beta}$, $n\geq 1$, or $E^{m\alpha+n\beta}$, $m,n\geq 2$. On the other hand, as explained above, if $\alpha+\beta\in\Phi(\mathfrak g)$, then
$4\alpha+n\beta$, $n\alpha+4\beta$, $n\geq 1$, and $m\alpha+n\beta$, $m,n\geq2$, cannot be roots, and the corresponding commutators vanish.
Summarizing, in the case
$\alpha+\beta\in\Phi(\mathfrak g)$, there are three possibilities. If neither $2\alpha+\beta$ nor $\alpha+2\beta$ belong to $\Phi(\mathfrak g)$, then
\be
\exp(\mu_\alpha E^\alpha)\exp(\mu_\beta E^\beta)= \exp\Big(\mu_\alpha E^\alpha+\mu_\beta E^\beta+{1\over2}\mu_\alpha\mu_\beta e_{\alpha \beta}E^{\alpha+\beta}\Big) \ ,
\label{eitheronezero}\ee
If $2\alpha+\beta\in \Phi(\mathfrak g)$, then
\begin{align}
&\exp(\mu_\alpha E^\alpha)\exp(\mu_\beta E^\beta)= \cr
& \exp\Big(\mu_\alpha E^\alpha+\mu_\beta E^\beta+{1\over2}\mu_\alpha\mu_\beta e_{\alpha \beta}  E^{\alpha+\beta}+{1\over12}\mu_\alpha^2\mu_\beta   e_{\alpha (\alpha+\beta)} e_{\alpha\beta}E^{2\alpha+\beta}\Big) \ ,
\label{eitherone}\end{align}
If $\alpha+2\beta\in \Phi(\mathfrak g)$, then
\begin{align}
&\exp(\mu_\alpha E^\alpha)\exp(\mu_\beta E^\beta) =\cr
&  \exp\Big(\mu_\alpha E^\alpha+\mu_\beta E^\beta+{1\over2}\mu_\alpha \mu_\beta e_{\alpha \beta}E^{\alpha+\beta}+{1\over12} \mu_\alpha \mu_\beta^2  e_{\beta (\alpha+\beta)} e_{\beta\alpha}E^{\alpha+2\beta}\Big) \ .
\label{eithertwo}\end{align}
The above results can be used as building blocks to solve more elaborated cases, such as
\be
\exp(E^\alpha)\exp(E^\beta)\exp(E^\gamma) \ .
\ee
Let us now consider $\exp({\mu_\alpha E^{\alpha}})\exp({\lambda_{\alpha}H^{\alpha}})
\exp({\mu_{-\alpha}E^{-\alpha}})$. Again, this is a particular case of the {\it type 4} commutator algebras \cite{Matone:2015xaa},
worked out in \cite{Matone:2015wxa}.
We have
\begin{align}
&\exp({\mu_\alpha E^{\alpha}})    \exp({\lambda_{\alpha}H^{\alpha}})
\exp({\mu_{-\alpha}E^{-\alpha}})= \cr\cr
& \exp\Big\{{\lambda_+-\lambda_-\over e^{-\lambda_-}-e^{-\lambda_+}}\Big[\mu_{+} E^\alpha+{1\over2}\Big(e^{-\lambda_+}+e^{-\lambda_-}-2\Big)H^\alpha+\mu_{-\alpha} E^{-\alpha}\Big]\Big\} \ ,
\label{lasoluzioneeXXX}\end{align}
where $\mu_{-\alpha},\lambda_{\alpha},\mu_\alpha\in\CC$, $\lambda_-+\lambda_+=\lambda_{\alpha}$, and
\be
e^{-\lambda_{\pm}}={1+e^{2\lambda_{\alpha}}+\mu_{-\alpha}\mu_\alpha\pm \sqrt{(1+e^{2\lambda_{\alpha}}+\mu_{-\alpha}\mu_\alpha)^2-4e^{2\lambda_{\alpha}}}\over2} \ .
\ee
Note that $\lambda_\alpha=0$ reproduces the closed form for $\exp({\mu_\alpha E^{\alpha}})
\exp({\mu_{-\alpha}E^{-\alpha}})$.

\noindent
The next case is $\exp({\mu_\alpha E^{\alpha}})\exp({\lambda_{\beta}H^{\beta}})
\exp({\mu_{\gamma}E^{\gamma}})$, with $\gamma\neq -\alpha$, and $\alpha+\gamma\notin\Phi({\mathfrak g})$, so that
\be
[E^\alpha,E^\gamma]=0 \ .
\ee
As we will discuss later, this corresponds to the {\it type 5} commutator algebras. However, it is instructive to directly derive the solution.
Consider the identity
\begin{align}
\exp({\mu_\alpha E^{\alpha}}) & \exp({\lambda_{\beta}H^{\beta}})
\exp({\mu_{\gamma}E^{\gamma}})= \cr\cr
& \exp({\mu_\alpha E^{\alpha}})\exp({\lambda_{\beta}^-H^{\beta}})\exp({\lambda_{\beta}^+H^{\beta}})
\exp({\mu_{\gamma}E^{\gamma}}) \ ,
\end{align}
where again
$\lambda_\beta^-+\lambda_\beta^+=\lambda_\beta$.
Then note that
\be
\exp({\mu_\alpha E^{\alpha}})\exp({\lambda_{\beta}^-H^{\beta}})=\exp\Big(\lambda_\beta^-H^\beta-{\lambda_\beta^-\mu_\alpha(\beta^\vee,\alpha)\over 1-e^{\lambda_\beta^-(\beta^\vee,\alpha)}} E^\alpha\Big) \ ,
\label{laprima}\ee
and
\be
\exp({\lambda_{\beta}^+H^{\beta}})\exp({\mu_\gamma E^{\gamma}})=\exp\Big(\lambda_\beta^+H^\beta+{\lambda_\beta^+\mu_\gamma(\beta^\vee,\gamma)\over 1-e^{-\lambda_\beta^+(\beta^\vee,\gamma)}} E^\gamma\Big) \ .
\label{laseconda}\ee
Imposing that the commutator between the exponents on the right hand side of (\ref{laprima}) and (\ref{laseconda}) vanishes
yields
\be
\mu_\gamma(\beta^\vee,\gamma)\Big(1-e^{\lambda_\beta^-(\beta^\vee,\alpha)}\Big)+\mu_\alpha(\beta^\vee,\alpha)\Big(1-e^{-\lambda_\beta^+(\beta^\vee,\gamma)}\Big)=0 \ .
\ee
It follows that
\begin{align}
\exp({\mu_\alpha E^{\alpha}}) & \exp({\lambda_{\beta}H^{\beta}})
\exp({\mu_{\gamma}E^{\gamma}})= \cr\cr
& \exp\Big(\lambda_\beta H^\beta-{\lambda_\beta^-\mu_\alpha(\beta^\vee,\alpha)\over 1-e^{\lambda_\beta^-(\beta^\vee,\alpha)}}
E^\alpha+{\lambda_\beta^+\mu_\gamma(\beta^\vee,\gamma)\over 1-e^{-\lambda_\beta^+(\beta^\vee,\gamma)}} E^\gamma\Big) \ .
\label{check}\end{align}
Such a case corresponds to the commutator algebras \cite{Matone:2015xaa}

\vspace{.8cm}

\noindent\underline{\it Type 5.}

{$$\Big[u\neq z, \, uz\neq0 \Big|m=-{uw\over z}, \, n=-vw\Big({1\over u}+{1\over z}\Big), \, p=-{vz\over u}, \, e=-{cw\over z}-{dv\over u}\Big|\Big]_{6}  $$}
$$
[X,Y]=uX+vY+cI  \qquad
[Y,Z]=wY+zZ+dI
$$
$$
[X,Z]=-{uw\over z}X-vw\Big({1\over u}+{1\over z}\Big)Y-{vz\over u}Z -\Big({cw\over z}+{dv\over u}\Big)I
$$
There are two equivalent solutions
\begin{align}
\alpha  &={v\over u} \nonumber \cr
\tilde u&=u-v-{uw\over z}  \nonumber \cr
\tilde v&=0  \nonumber \cr
\tilde c &= \Big(1-{v\over u}-{w\over z}\Big)d \nonumber
\end{align}
and
\begin{align}
\alpha  &=1-{w\over z} \nonumber \cr
\tilde u&=0  \nonumber \cr
\tilde v&=z-w-{vz\over u}  \nonumber \cr
\tilde c &= \Big(1-{v\over u}-{w\over z}\Big)d \nonumber
\end{align}
In the case at hand, the non-vanishing parameters are only $u=-\lambda_\beta(\beta^\vee,\alpha)$ and $z=\lambda_\beta(\beta^\vee,\gamma)$. It is
immediate to check that using one of the two solutions one gets (\ref{check}).

\section{Generalization of the algorithm by iteration}

In this section we iterate the algorithm introduced in \cite{Matone:2015wxa} to extend the Van-Brut and Visser original formula.
The problem then is to find, in more general cases, the closed form $W$, defined by
\be
\exp(X)\exp(Z)=\exp(W) \ .
\label{tricksuno}\ee
A first result in
this direction was already obtained in \cite{Matone:2015wxa}. The point is to start with the BCH problem for the product of the three exponentials in (\ref{BCHproblem}), choosing
$Y$ in such a way that (\ref{riecco}) and (\ref{richiamare}) are satisfied. Next, one reproduces the steps of the algorithm
leading to (\ref{llasol}) where $Y$ is replaced by $\lambda Y$ and then considers
\be
\exp(X)\exp(Z)=\lim_{\lambda\to 0}\exp(X)\exp(\lambda Y)\exp(Z) \ .
\label{tricksdue}\ee
Note that the relations (\ref{riecco}) and (\ref{richiamare}) satisfied by $X$, $Y$ and $Z$ can be also expressed
in terms of $X$, $\lambda Y$ and $Z$ with $u$, $c$, $z$, $d$ and $Y$ multiplied by $\lambda$, whereas $n$ is replaced
by $n/\lambda$
\begin{align}
[X,Z]= & mX+\Big({n\over\lambda}\Big) (\lambda Y)+p Z + eI \ , \cr\cr
[X,(\lambda Y)] & =(\lambda u) X+v(\lambda Y)+(\lambda c) I \ , \cr\cr
[(\lambda Y),Z] &=w(\lambda Y)+(\lambda z) Z+(\lambda d) I \ .
\label{eccod}\end{align}
It follows that the expression on the right hand side of (\ref{tricksdue}), before taking the $\lambda\to0$ limit,
 is given by (\ref{llasol}) with $u$, $c$, $z$, $d$ and $Y$ multiplied by $\lambda$
\begin{align}
\tilde X & =g_{\alpha}(\lambda u,v)X+h_\alpha(\lambda u,v)(\lambda Y)+l_{\alpha}(\lambda u,v)(\lambda c) I \ , \cr
\tilde Y & = h_\beta(\lambda z,w)(\lambda Y)+g_\beta(\lambda z,w) Z+l_\beta(\lambda z,w)(\lambda d)I \ .
\end{align}
It then follows by (\ref{llasol})
\begin{equation}
\exp(X) \exp(Z)=\lim_{\lambda\to0} \exp({\tilde X+\tilde Y+f(\tilde u,\tilde v)[\tilde X,\tilde Y]}) \ .
\label{llasolbisg}\end{equation}
On the other hand, in the $\lambda\to0$ limit we get $\tilde X= X$, $\tilde Y= Z$ and, by (\ref{utildevtilde}),
$\tilde u=m$ and $\tilde v=p$. Hence
\be
\exp(X)\exp(Z)=\exp(X+Z+f(m,p)[X,Z]) \ ,
\label{tricksdugento}\ee
showing that the Van-Brut and Visser formula (\ref{bbbvvv}) extends to the case in which there exists $Y$ such that
the weaker conditions (\ref{eccod}) are satisfied.
Note that in the $\lambda\to 0$ limit the linear system
for $e$, $m$, $n$ and $p$ is again (\ref{AsystemINTRO}).

\noindent The above results can be summarized by the following Lemma.
\\

\noindent {\it Lemma 1.} If
\be
[X,Z]=mX+nY+pZ+eI \ ,
\ee
then the Van-Brut and Visser formula
\be
\exp(X)\exp(Z)=\exp(X+Z+f(m,p)[X,Z]) \ ,
\ee
holds if $Y$ satisfies the conditions
\be
[X,Y]=uX+vY+cI \ , \qquad  [Y,Z]=wY+z Z+dI \ .
\ee

\noindent
It is clear that the above construction can be iterated to generalize the algorithm leading to (\ref{llasol}). Namely, one can repeat the steps of the algorithm,
where now the Van-Brut and Visser condition is replaced by the weaker one given by Lemma 1.
In particular, the first step of the algorithm is again to consider the decomposition
\begin{equation}
\exp(X) \exp(Y) \exp(Z)= \exp(X)\exp({\alpha Y}) \exp({\beta Y}) \exp(Z) \ ,
\label{decomposizioneTTTD}\end{equation}
with $\alpha+\beta=1$.
However, now each one of the two conditions (\ref{riecco}) should be replaced by the weaker conditions in Lemma 1. Namely,
if there exist $Y'$ and $Y''$ satisfying the conditions
\begin{align}
[X,Y] & = a_XX +a_YY+a_{Y'}Y'+aI \ , \cr
[X,Y'] & =b_X X + b_{Y'} Y'+bI \ , \cr
[Y',Y] & =c_{Y'}Y'+c_YY+cI \ ,
\label{lafirst}\end{align}
and
\begin{align}
[Y,Z] & =d_YY +d_{Y''}Y''+d_ZZ+dI \ , \cr
[Y,Y''] & =e_YY + e_{Y''} Y''+eI \ , \cr
[Y'',Z] & =f_{Y''}Y''+f_ZZ+fI \ ,
\label{lasecond}\end{align}
then one can still apply the algorithm leading to (\ref{llasol}). The expression of $\exp(X) \exp(Y) \exp(Z)$ is still
the one in (\ref{llasol}), but now there are good news, namely  the condition (\ref{riecco}) is replaced by the weaker conditions (\ref{lafirst}) and (\ref{lasecond}).
Furthermore, the conditions to be imposed on
\begin{align}
\tilde X &:=\ln(e^Xe^{\alpha Y})= X+\alpha Y+f(\alpha a_X,a_Y)[X,\alpha Y] \ , \cr
\tilde Y &:=\ln( e^{\beta Y}e^Z ) = \beta Y + Z + f(d_Y,\beta d_Z)[\beta Y,Z] \ ,
\label{conicommutatori}\end{align}
that we rewrite as
\begin{align}
\tilde X &= g_{\alpha}(a_X,a_Y)X+h_{\alpha}(a_X,a_Y)Y+l_{\alpha}(a_X,a_Y)(a_{Y'}Y'+aI) \ , \cr
\tilde Y &= h_{\beta}(d_Y,d_Z)Y+g_{\beta}(d_Y,d_Z)Z+l_{\beta}(d_Y,d_Z)(d_{Y''}Y''+dI) \ ,
\label{colon0b}\end{align}
are weaker with respect to (\ref{labella}). Namely,
according to Lemma 1., we now impose that there exists $\tilde Y'$ such that
\begin{align}
[\tilde X,\tilde Y] & = a_{\tilde X}\tilde X +a_{\tilde Y}\tilde Y+a_{\tilde Y'}\tilde Y'+\tilde aI \ , \cr
[\tilde X,\tilde Y'] & =b_{\tilde X} \tilde X + b_{\tilde Y'} \tilde Y'+\tilde bI \ , \cr
[\tilde Y',\tilde Y] & =c_{\tilde Y'}\tilde Y'+c_{\tilde Y}\tilde Y+\tilde cI \ .
\label{lathird}\end{align}
By (\ref{lafirst}), (\ref{lasecond}), (\ref{colon0b}) and (\ref{lathird}) it follows that now $X$ and $Z$ satisfy the weaker condition
\be
[X,Z]=mX+nY+pY'+qY''+r\tilde Y'+sZ+tI \ .
\ee
One then can apply the rest of the algorithm and classify all the possible algebras leading to the closed BCH formula, in a similar way to the classification developed in \cite{Matone:2015xaa}.

\noindent The next step is to investigate again (\ref{tricksdue}), but now using the weaker conditions (\ref{lafirst}), (\ref{lasecond}) and (\ref{lathird}). This is the content
of Lemma 2. analogous to Lemma 1. but with the previous weaker conditions on the commutators.
\\

\noindent {\it Lemma 2.} If
\be
[X,Z]=mX+nY+pY'+qY''+r\tilde Y'+sZ+tI\ ,
\ee
then the Van-Brut and Visser formula
\be
\exp(X)\exp(Z)=\exp(X+Z+f(m,p)[X,Z]) \ ,
\ee
holds if $Y$, $Y'$, $Y''$ and $\tilde Y'$ satisfy the conditions  (\ref{lafirst}), (\ref{lasecond}) and (\ref{lathird}).
\\

\noindent
Such an algorithm can be iterated indefinitely by applying
again the original algorithm to (\ref{decomposizioneTTTD}) but now using Lemma 2., applying the result to (\ref{tricksdue})
will lead to Lemma 3., with weaker conditions with respect to Lemma 2., etc..
The full consequences of such a construction are under investigation.

\section*{Acknowledgements} It is a pleasure to thank  Andrea Bonfiglioli, Pieralberto Marchetti,  Paolo Pasti,
 Dmitri Sorokin, Roberto Volpato and Roberto Zucchini for interesting
discussions.

\end{document}